\documentclass[12pt]{article}

 \usepackage{amssymb}

 \newcommand{\nn}{\nonumber}
 \newcommand{\p}[1]{(\ref{#1})}
 \newcommand{\be}{\begin{equation}}
 \newcommand{\ee}{\end{equation}}
 \newcommand{\bea}{\begin{eqnarray}}
 \newcommand{\eea}{\end{eqnarray}}
 \def\tr{{\rm tr}}

 \def\theequation{\arabic{section}.\arabic{equation}}
 \def\tfrac#1#2{{\textstyle {#1 \over #2}}}
 \def\dfrac#1#2{{\displaystyle {#1 \over #2}}}
 \def\dint{\displaystyle \int }

 \def\cN{{\cal N}}
 \def\beq{\begin{eqnarray}}
 \def\eeq{\end{eqnarray}}
 \newcommand{\m}{d\zeta(^{33}_{11})du\,}

\begin{document}
 \begin{titlepage}
 \begin{flushright}
 \tt hep-th/0403053
 \end{flushright}
 \vspace{1cm}
 \begin{center}
 {\Large\bf
Scale Invariant Low-Energy Effective Action }

\vspace{0.2cm}
{\Large\bf in $\cN=3$ SYM Theory}
 \\[0.5cm]
 {\bf
 I.L. Buchbinder $^+$\footnote{joseph@tspu.edu.ru},
 E.A. Ivanov $^\dag$\footnote{eivanov@thsun1.jinr.ru},
 I.B. Samsonov $^{*}$\footnote{samsonov@phys.tsu.ru},
 B.M. Zupnik $^\dag$\footnote{zupnik@thsun1.jinr.ru}}\\[3mm]
 {\it $^+$ Department of Theoretical Physics, Tomsk State Pedagogical
 University,\\ Tomsk 634041, Russia\\[2mm]
 $^\dag$ Bogolubov Laboratory of Theoretical Physics, Joint Institute for
 Nuclear Research, Dubna, 141980 Moscow Region, Russia\\[2mm]
 $^*$ Laboratory of Mathematical Physics, Tomsk Polytechnic
 University,\\ 10 Lenin Ave, Tomsk 634034, Russia
 }
 \\[0.8cm]
\bf Abstract
\end{center}
Using the harmonic superspace approach we study the problem of
low-energy effective action in $\cN=3$ SYM theory.
The candidate effective action is a scale and $\gamma_5$-invariant
functional in full $\cN=3$ superspace built out of $\cN=3$ off-shell superfield strengths.
This action is constructed as $\cN=3$ superfield generalization of
$F^4/\phi^4$ component term which is leading in the low-energy effective action
and is simultaneously the first nontrivial term in scale invariant
Born-Infeld action. All higher-order terms in the scale invariant Born-Infeld action are
also shown to admit an off-shell superfield completion in $\cN=3$ harmonic superspace.
\end{titlepage}
\setcounter{footnote}{0}

 \section{Introduction}
Field theories with extended supersymmetry play an important role in modern
high-energy physics due to their beautiful properties and intertwining relationships
with string theory. A vastly known example is $\cN=4$ super Yang-Mills (SYM) theory.
Its geometric and quantum aspects were a subject of numerous studies.
It was found that the symmetries of this theory (unified into $\cN=4$ superconformal
symmetry) are so rich  that many its remarkable quantum properties can be proved solely
on the symmetry grounds. E.g. the ultraviolet finiteness
of quantum $\cN=4$ SYM theory can be attributed to the exact superconformal
invariance of the latter. Some low-energy contributions to the quantum effective action in the $\cN=2$
gauge multiplet sector can be exactly found on the basis of smaller scale
and $\gamma_5$ invariances \cite{Dine} (see also review \cite{bbiko}).
Further invoking the hidden (on-shell) $\cN=4$ supersymmetry allows one
to extend the analysis of \cite{Dine}
to the hypermultiplet sector and to reveal the complete structure of one-loop
effective action on the Coulomb branch \cite{BuchIvanov}. Many proofs and
computations are essentially simplified with using the off-shell approach of
$\cN=2$ harmonic superspace (HSS) \cite{GIK1,IvanovBook} which makes manifest two
of four underlying Poincar\'e supersymmetries of the $\cN=4$ SYM theory.

In this paper we undertake some steps towards studying, along similar lines,
the $\cN=3$ SYM theory in $\cN=3$ HSS \cite{Harm1}.
This superspace is a generalization of $\cN=2$ HSS,
and provides a formulation of $\cN=3$ gauge theory in terms of unconstrained off-shell $\cN=3$
superfields. Such a model, like $\cN=4$ SYM, is
known to be finite \cite{Delduc} and superconformally invariant \cite{Harm2}.
Actually, $\cN=3$ SYM model describes the dynamics of the same multiplet of
physical field as the $\cN=4$ one (see e.g. book \cite{book2} for a review),
so these theories are different only off shell. Thus $\cN=3$ SYM theory in
$\cN=3$ HSS can be regarded as a superfield formulation of
$\cN=4$ SYM theory with bigger number of off-shell supersymmetries compared
to the more familiar  $\cN=2$ superfield formulation of $\cN=4$ SYM theory.
Recently, it was argued \cite{hs} that
the $\cN=3$ superfield formulation of $\cN=4$ SYM theory in fact
reveals the maximally attainable  number of off-shell supersymmetries,
and there is no way to continue off shell the entire $\cN=4$ supersymmetry. This shows the
urgency and importance of studying quantum properties of the $\cN=3$ SYM model.
So far they were not explored  at all, and the structure of the quantum effective
$\cN=3$ harmonic superfield action of $\cN=3$ SYM theory remains to be disclosed.
The language of $\cN=3$ HSS is capable to greatly simplify quantum calculations,
e.g. due to the fact that the interaction superfield Lagrangian of $\cN=3$ SYM
theory includes only one or two vertices (in the first- or second-order
formalisms \cite{Harm1,IvanovBook}), which should be contrasted with an infinite tower of
vertices in the $\cN =2$ HSS formulation of $\cN=4$ SYM theory.
As another evidence in favour of usefulness of the $\cN=3$ HSS
approach, it was recently shown \cite{Grassi} that $\cN=3$ SYM theory
in harmonic superspace is naturally generated from superstring theory.

It is well known \cite{Dine,bbiko,BuchN4} that the leading term in the
 low-energy effective action of $\cN=4$ SYM model in the sector of $\cN=2$
 vector multiplet has the form $\int d^4x\frac{F^2\bar
 F^2}{(\bar\varphi\varphi)^2}$, where $F^{\alpha\beta},\bar
 F^{\dot\alpha\dot\beta}$ are the Abelian field strengths and $\varphi$ is
 the scalar field belonging to $\cN=2$ vector multiplet. Complete
 $\cN=4$ supersymmetric generalization of the low-energy effective action
 including both vector fields and hypermultiplets in the $\cN=2$ HSS approach
 was given in \cite{BuchIvanov}.\footnote{Note also recent papers
 \cite{general} devoted to discussion of general structure of the
 effective actions within the $\cN=2$ HSS approach.} The leading bosonic component
of this action is given by
 \be
 \int d^4x \frac{F^2\bar F^2}{(\bar \phi_i\phi^i)^2}
 \label{F4}
 \ee
 where $\phi^i$ is a complex SU(3) triplet of scalar fields.\footnote{In \cite{BuchIvanov}
 these fields were arranged into a real self-dual 6-plet of SU(4).}
 Since the models of $\cN=3$ and $\cN=4$ SYM are equivalent on-shell, we
 expect that the term (\ref{F4}) is also leading in the effective action
 of $\cN=3$ SYM model. Therefore it is interesting and important to find $\cN=3$
 superfield action which would reproduce the expression (\ref{F4}) in the
 component expansion. One can expect that such an action presents the low-energy
 quantum effective action of $\cN=3$ SYM model.

 An important step in understanding the possible structure of the
 effective action in $\cN=3$ gauge theory was the construction of $\cN=3$
 supersymmetric Born-Infeld (BI) action  in $\cN=3$ HSS \cite{Zupnik01}.
 However, this BI action contains a dimensionful parameter, hence it is not scale
 invariant and cannot reproduce scalar fields in the denominator of
 the expression (\ref{F4}). It is natural to assert that the low-energy effective
 action of $\cN=3$ gauge theory corresponds to a scale invariant
 generalization of the $\cN=3$ BI theory. In this work we propose a
 possible form of such a superfield functional which meets the requirements of
 $\cN=3$ supersymmetry, as well as gauge, scale and $\gamma_5$ invariances, and
 gives rise to the term (\ref{F4}) after going to components.

 The paper is organized as follows. In section 2, just to fix our notation, we
 overview some facts about $\cN=3$ BI and SYM theories in $\cN=3$ HSS.
 In section 3 we propose a possible HSS form of a scale invariant
 $\cN=3$ supersymmetric functional reproducing the leading term
 in the low-energy effective action of $\cN=3$ SYM theory and
 study the component structure of this functional.
 In section 4 we present $\cN=3$ supersymmetric scale invariant
 completion of all higher terms in BI action, starting with
 $F^8/\phi^{12}$. Section 5 contains concluding remarks and discussions.
 In the Appendix we collect useful formulas of the $\cN=3$ HSS formalism.

\setcounter{equation}0
 \section{$\cN=3$ Supersymmetric Yang-Mills and Born-Infeld actions}

The $\cN=3$ HSS was introduced in ref. \cite{Harm1} to construct, for the first time,
an off-shell superfield formulation of $\cN=3$ SYM model. The basics of the harmonic superspace
method are exposed in book \cite{IvanovBook}. Throughout this paper we
follow the conventions of ref. \cite{Zupnik01}. Some basic formulas of the
$\cN=3$ HSS approach are collected in the Appendix.

The classical HSS action of $\cN=3$ SYM model is
\begin{eqnarray}
S^{\cN=3}_{SYM}[V]&=&S_2[V]+S_{int}[V],\nonumber\\
S_2[V]&=&-\frac14\tr\dint\m[V^2_3D^1_3V^1_2+\frac12(D^1_2V^2_3-D^2_3V^1_2)^2]\,,
 \label{S2_}\\
S_{int}[V]&=&-\frac14\tr\dint\m \{\, i
(D^1_2V^2_3-D^2_3V^1_2)[V^1_2,V^2_3]-\frac12[V^1_2,V^2_3][V^1_2,V^2_3]\}.
 \label{e5}
\end{eqnarray}
The piece $S_2[V]$ is the free action and $S_{int}[V]$ represents the
interaction. We will consider only the Abelian case when the interaction is absent.

The prepotentials $V^1_2$, $V^2_3$ are analytic superfields and the
integration in (\ref{e5}) is performed over analytic superspace
(see eq. (\ref{gme2}) for the definition of the analytic superspace integration measure).
Physical bosonic component fields are contained in the prepotentials as \cite{Harm1}
 \be
 \begin{array}{rl}
 V^2_3=&[(\bar\theta^1\bar\theta^2)u^2_k-(\bar\theta^2)^2u^1_k]\phi^k
 +\theta_3^\alpha\bar\theta^{2\dot\alpha}A_{\alpha\dot\alpha}
 -i\theta_2^\alpha\theta_3^\beta(\bar\theta^2)^2H_{\alpha\beta}\\
 &+{\rm spinors\ and\ auxiliary\ fields}\,,\\
 V^1_2=&-\widetilde{(V^2_3)}=-[(\theta_2\theta_3)\bar u_2^k-(\theta_2)^2\bar
 u_3^k]\bar\phi_k+\theta_2^\alpha\bar\theta^{1\dot\alpha}A_{\alpha\dot\alpha}
 +i(\theta_2)^2\bar\theta^{1\dot\alpha}\bar\theta^{2\dot\beta}
  \bar H_{\dot\alpha\dot\beta}\\
 &+{\rm spinors\ and\ auxiliary\ fields}\,.
 \end{array}
 \label{e4}
 \ee
 Here, $\phi^i,\bar\phi_i$ are complex scalar fields,
 $A_{\alpha\dot\alpha}$ is a vector gauge field, $H_{\alpha\beta}$,
 $\bar H_{\dot\alpha\dot\beta}$ are the auxiliary fields which ensure
 the correct structure of the gauge field sector of the theory \cite{Zupnik01}.

 The component form of the action $S_2$ in the SU(3) singlet gauge field sector
 is \cite{Zupnik01}
 \be
 S_2=\int d^4x[V^2+\bar V^2-2(\bar V\bar F+ VF)+\frac12(F^2+\bar F^2)]
 \label{S2}
 \ee
 where
 \be
 \begin{array}l
 V_{\alpha\beta}=\frac14(H_{\alpha\beta}+F_{\alpha\beta}),\qquad
 \bar V_{\dot\alpha\dot\beta}=\frac14(\bar H_{\dot\alpha\dot\beta}+
  \bar F_{\dot\alpha\dot\beta}),\\
 F^2=F^{\alpha\beta}F_{\alpha\beta},\quad
 V^2=V^{\alpha\beta}V_{\alpha\beta},\quad
 FV=F^{\alpha\beta}V_{\alpha\beta},\\
 F_{\alpha\beta}=\frac i4(\sigma_{mn})_{\alpha\beta}(\partial_mA_n-\partial_n
 A_m),\qquad
 \bar
 F_{\dot\alpha\dot\beta}=-\frac
i4(\tilde\sigma_{mn})_{\dot\alpha\dot\beta}(\partial_mA_n-\partial_nA_m).
 \end{array}
 \ee
 The auxiliary fields $V_{\alpha\beta}$, $\bar V_{\dot\alpha\dot\beta}$ can be
 eliminated by their algebraic classical equations of motion
 \be
 V_{\alpha\beta}=F_{\alpha\beta},\qquad
 \bar V_{\dot\alpha\dot\beta}=\bar F_{\dot\alpha\dot\beta}.
 \label{aux}
 \ee
 As a result, the free classical action (\ref{S2}) takes the form of the usual
 Maxwell action
 \be
 S_2=-\frac12\int d^4x(F^2+\bar F^2)\,.
 \label{maxwell}
 \ee

The higher order terms of the superfield ${\cal N}=3$ BI-action
can be constructed out of $\cN=3$ superfield strengths which are
expressed through prepotentials as
\be
\begin{array}{ll}
W_{23}=\frac14\bar D_{3\dot\alpha}\bar D_3^{\dot\alpha}V^3_2\,,\qquad&
\bar W^{12}=-\frac14D^{1\alpha}D^1_\alpha V^2_1\,,\\
W_{12}=D^3_1 W_{23},& \bar W^{23}=-D^3_1\bar W^{12}\,,\\
W_{13}=-D^2_1W_{23},& \bar W^{13}=D^3_2 \bar W^{12}\,.
\end{array}
\label{str}
\ee
Here $V^2_1$, $V^3_2$ are non-analytic prepotentials which are the
solutions of zero-curvature equations \cite{Zupnik01}
\be
D^2_1 V^1_2= D^1_2 V^2_1\,,\qquad
D^3_2 V^2_3= D^2_3 V^3_2\,.
\label{zero-curv}
\ee
 The superfields (\ref{str}) have the following component structure
 in the sector of physical bosons
 \cite{Ferrara}
 \be
 \begin{array}{rl}
 W_{23}=&u^1_i\phi^i(x_{A+})+4i\theta_2^\alpha\theta_3^\beta
 V_{\alpha\beta}(x_{A+})
 +{\rm spinors\ and\ auxiliary\ fields}\,,\\
 \bar W^{12}=&\bar u_3^i\bar\phi_i(x_{A-})
 +4i\bar\theta^{1\dot\alpha}\bar\theta^{2\dot\beta}
 \bar V_{\dot\alpha\dot\beta}(x_{A-})
 +{\rm spinors\ and\ auxiliary\ fields}
 \end{array}
 \label{e10}
 \ee
where $x_{A\pm}^{\alpha\dot\alpha}=x_A^{\alpha\dot\alpha}\pm
2i\theta_2^\alpha\bar\theta^{2\dot\alpha}$.

The $\cN=3$ supersymmetric Born-Infeld action \cite{Zupnik01}
can be represented in the following form:
\begin{eqnarray}
S_{BI}^{\cN=3}&=&S_2+S_E,\nonumber\\
S_E=\sum\limits_{n=1}^\infty S_{4n}&=&\frac1{32X^2}\dint\m (\bar W^{12}W_{23})^2\hat E(A/X^4)
\label{N3BI}
\end{eqnarray}
where $X=f^{-1}$ is a coupling constant of the mass dimension 2.
By definition,
the composite analytic superfield contains the 4-th degree of the auxiliary
fields
\be
A=\frac1{2^{10}}(D^1)^2(\bar D_3)^2[
D^{2\alpha}W_{12}D^2_\alpha W_{12}\bar D_{2\dot\alpha}\bar W^{23}\bar D_2^{\dot\alpha}\bar
W^{23}]=V^2\bar V^2+\ldots\,.
\label{A}
\ee
The superfield function $\hat E(A)$ is connected with the self-interaction
function of the auxiliary fields in the BI theory $E(a)=E(V^2\bar V^2)$
 \cite{Zupnik01}
\be
\hat E(a)=\frac2a E(a)=\frac4a[2t^2(a)+3t(a)+1],\qquad
t^4+t^3-\frac14a=0,\qquad
t|_{a=0}=-1.
\label{EA}
\ee
The series expansion for the function $\hat E(a)$ starts with
\be
\hat E(a)=1-\frac a4+\frac{3a^2}{16}+\ldots
\ee
where the 1-st term corresponds to the 4-th order interaction $S_4$.
In the gauge field sector the action (\ref{N3BI}) yields the
standard bosonic BI action \cite{Zupnik01}
\begin{eqnarray}
S_{BI}&=&\int d^4x\left[-\frac12(F^2+\bar F^2)+\frac12 \frac{F^2\bar F^2}{X^2}-
\frac14\frac{F^2 \bar F^2(F^2+\bar F^2)}{X^4} \right. \nn \\
&&\left.  +\,\frac18\frac{F^2\bar F^2(3F^2\bar F^2+F^4+\bar
F^4)}{X^6}+\ldots \right]\nonumber\\
&=& X^2\,\int d^4x\left[\sqrt{-\det(\eta_{mn}+F_{mn}/X)}-1\right]
\label{BI}
\end{eqnarray}
where $F_{nm}=\partial_n A_m-\partial_m A_n$ is the Maxwell
field strength.

We emphasize the fact that the action $S_E$ (\ref{N3BI}), as well as its
component expansion (\ref{BI}), contains the coupling constant $X$ of mass dimension
$2$. Therefore this action is not scale invariant.

Let us single out the quartic superfield part in the action (\ref{N3BI})
\be
S_4=\frac1{32}\dint\m \frac{(\bar W^{12}W_{23})^2}{X^2}\,.
\label{S4_}
\ee
This action produces the first nontrivial term of the BI interaction
$E(V^2\bar V^2)$
\be
\frac12\int d^4x \frac{V^2\bar V^2}{X^2}
\label{F44}
\ee
and makes the corresponding contributions to the BI action (\ref{BI}), starting from
the $F^2\bar F^2$ term (in particular, it uniquely fixes the coefficient before the 6-th order
term). Note that the next, 8-th order superfield term in \p{N3BI} produces
the unique term $\sim (V^4\bar V^4)$ in the lagrangian of
tensor auxiliary fields and, nevertheless, affects the whole infinite sequence of terms
$F^{4(1+k)}\bar F^{4(1+k)}\,,$ $(k\geq 0)$ in the effective Maxwell field strength action
after eliminating the auxiliary fields by their equations of motion. The correct BI action
is restored only after accounting all superfield terms in \p{N3BI}.

\setcounter{equation}0
\section{Scale and R-invariant $\cN=3$ superfield low-energy effective action}
In this Section we construct a manifestly $\cN=3$ supersymmetric low-energy
effective action containing the term $F^4/{\phi}^4$ in the bosonic sector.

$\cN=3$ SYM theory is known to be a superconformal field theory
\cite{Harm2}, like the $\cN=4$ SYM one. Moreover, both these models
describe the dynamics of the same multiplet of physical fields and
therefore are on-shell equivalent \cite{book2}. The
effective action of $\cN=3$ SYM model should be scale invariant.
The mutually commuting transformations of dilatations (scale invariance) and $\gamma_5$-symmetry
(R-symmetry) act on the coordinates of harmonic superspace and superfield strengths as follows
\bea
&&\delta x^m_A=ax^m_A,\quad\delta\theta^\alpha_I=\frac12(a+ib)\theta^\alpha_I,
\quad\delta\bar\theta^{I\dot\alpha}=\frac12(a-ib)\bar\theta^{I\dot\alpha}
\nonumber\\
&&\delta W_{IJ}=(-a+ib)W_{IJ},\quad \delta\bar W^{IJ}=-(a+ib)\bar W^{IJ}.
\eea
We expect that a scale and $\gamma_5$-invariant generalization of the action
(\ref{S4_}) should correspond to the low-energy effective action of $\cN=3$
SYM model. In components such an action should reproduce
the scale and $\gamma_5$-invariant generalization of (\ref{F44}), that is (\ref{F4}).
Note that exactly this term is leading in the low-energy effective action
of $\cN=4$ SYM model \cite{Dine,BuchN4}.
Thus we wish to construct a generalization of the action
(\ref{S4_}) which would bear the scale- and $\gamma_5$-invariances.
Note that the effective ${\cal N}=3$ action of ref. \cite{Zupnik01}
is $\gamma_5$-invariant, so we shall discuss its scale invariant generalization.

To pass from (\ref{F44}) to the scale invariant component action (\ref{F4}),
one should replace the dimensionful constant $X$ by
the function of scalar fields $(\phi^i\bar\phi_i)$. Therefore, to obtain
a scale invariant generalization of the superfield action (\ref{S4_}) we have to replace the
constant $X$ by some superfield expression having the
same dimension and containing $\phi^i\bar\phi_i$ as the lowest component.
The suitable expression is
\be
\bar W^{IJ}W_{IJ}=\bar W^{12}W_{12}+\bar W^{23}W_{23}+\bar W^{13}W_{13}\,.
\label{WW}
\ee
Indeed, the component expansion of this superfield starts with the scalars
(see \cite{Ferrara} for details)
\be
\bar W^{IJ}W_{IJ}|_{\theta=\bar\theta=0}=\phi^i\bar\phi_i\,.
\label{WW|}
\ee
It is worth noting that this superfield, as well as the basic
analytic superfields $W_{23}$ and $\bar W^{12}$, are covariant with respect
to the whole superconformal ${\cal N}=3$ supergroup \cite{IvanovBook}. However,
we do not study the superconformal properties of the effective action here.

The expression (\ref{WW}) cannot be naively inserted into the
integral in (\ref{S4_}) in place of the constant $X$. The point is that
the superfield $(\bar W^{IJ}W_{IJ})$ is not analytic since the superfield strengths
$\bar W^{23},\bar W^{13},W_{12},W_{13}$ are not analytic, while the
integration in (\ref{S4_}) goes over the analytic superspace. Therefore we
have to rewrite the action (\ref{S4_}) in full $\cN=3$ HSS and then to insert
$\bar W^{IJ}W_{IJ}$ into the integral.

The action (\ref{S4_}) in the full $\cN=3$ HSS is written as
\be
S_4=\frac{1}{32}\int d^4x d^{12}\theta du \frac1{X^2}\left[\frac{(\bar
D_1)^2}{4\square}(W_{23})^2\right]
\left[\frac{(D^3)^2}{4\square}(\bar W^{12})^2\right].
\label{S4full}
\ee
To check that the actions (\ref{S4_}) and (\ref{S4full}) are actually identical to each other,
one should express the integration measure of the full $\cN=3$ superspace through the analytic one
\be
d^4x d^{12}\theta=d\zeta(^{33}_{11})\,\frac{1}{4}(D^1)^2\, \frac{1}{4}(\bar D_3)^2\,,
\label{measures}
\ee
and then apply the anticommutation relations
\be
\{D^1_\alpha,\bar D_{1\dot\alpha}\}=\{D^3_\alpha,\bar D_{3\dot\alpha}\}
=-2i(\sigma^m)_{\alpha\dot\alpha}\partial_m .
\label{commut}
\ee

Replacing the constant $X$ by the superfield $\bar W^{IJ}W_{IJ}$
in (\ref{S4full}), we arrive at the action
\be
S_{4}^{scale-inv}=\alpha\int d^4xd^{12}\theta du\frac 1{(\bar W^{IJ}W_{IJ})^2}
\left[\frac{(\bar D_1)^2}{4\square}(W_{23})^2\right]
\left[\frac{(D^3)^2}{4\square}(\bar W^{12})^2\right]
\label{Seff}
\ee
where $\alpha$ is some dimensionless constant.
This constant can be fixed from a straightforward calculation of low-energy
effective action in the framework of quantum field theory.
Besides, if one assumes e.g. that the term (\ref{Seff}) is a part of supersymmetric
BI action, the constant $\alpha$ can be fixed from the requirement that
the (\ref{Seff}) reproduces the term $F^4/\phi^4$ in the
scale invariant version of the expression (\ref{BI}) with the correct coefficient.
Since the action (\ref{Seff}) includes no any dimensional constants, it is
scale invariant. It can be checked to be $\gamma_5$-invariant as well.

Let us study the component structure of the action (\ref{Seff}).
Note that the superfield strengths entering the action contain a multiplet
of physical fields as well as an infinite number of auxiliary fields.
We are interested in the component structure of the action (\ref{Seff}) in
the sector of scalar and vector physical fields. For this purpose
we neglect all the derivatives on scalar fields and Maxwell field strength.
Such an approximation is sufficient for retrieving the term $F^4/\phi^4$
while going to components. Therefore we use the following ansatz for the superfield strengths
\be
\begin{array}{ll}
\hat{\bar W}{}^{12}=\bar\phi_3+\bar\omega^{12}\,, & \hat W_{12}=\phi^3+\omega_{12}\,,\\
\hat{\bar W}{}^{23}=\bar\phi_1+\bar\omega^{23}\,, & \hat W_{23}=\phi^1+\omega_{23}\,,\\
\hat{\bar W}{}^{13}=-\bar\phi_2+\bar\omega^{13}\,, & \hat W_{13}=-\phi^2+\omega_{13}
\end{array}
\label{str-comp}
\ee
where
\be
\begin{array}{ll}
\bar\phi_I=\bar u_I^i\bar\phi_i,& \phi^I=u^I_i\phi^i\,,\\
\bar\omega^{IJ}=4i\bar\theta^{I\dot\alpha}\bar\theta^{J\dot\beta}
 \bar V_{\dot\alpha\dot\beta}\,,&
\omega_{IJ}=4i\theta_I^\alpha\theta_J^\beta V_{\alpha\beta}\,.
\end{array}
\label{str-comp1}
\ee
Here $V_{\alpha\beta}, \bar V_{\dot\alpha\dot\beta}$ are auxiliary tensor fields
which have the same properties as the Maxwell field strengths $F_{\alpha\beta}, \bar
F_{\dot\alpha\dot\beta}$ (the fields
$V_{\alpha\beta}, \bar V_{\dot\alpha\dot\beta}$ are eventually expressed through
$F_{\alpha\beta}, \bar F_{\dot\alpha\dot\beta}$). The symbol ``hat'' below
indicates that we consider only scalar and vector bosonic fields and discard any
auxiliary fields, except for the SU(3) singlet tensor ones just defined.
With the ansatz (\ref{str-comp}) and (\ref{str-comp1}), we have
\be
(\bar D_1)^2(\hat W_{23})^2=4(\theta_1)^2\square (\hat W_{23})^2,\qquad
(D^3)^2(\hat{\bar W}{}^{12})^2=4(\bar\theta^3)^2\square(\hat{\bar W}{}^{12})^2.
\label{d1}
\ee
Therefore, when we consider only the scalar and vector fields, the action
(\ref{Seff}) contains only local terms
\be
\hat S_{4}^{scale-inv}=\alpha\int d^4x d^{12}\theta du \frac{(\theta_1)^2(\bar\theta^3)^2}{
(\hat{\bar W}{}^{IJ}\hat W_{IJ})^2}(\hat{\bar W}{}^{12}\hat W_{23})^2.
\label{S1}
\ee

To find the component structure of the action (\ref{S1}) we need the
following expansions:
\be
\begin{array}{rl}
(\hat{\bar W}{}^{12}\hat W_{23})^2=&(\bar\phi_3\phi^1)^2+(\bar\phi_2\omega_{23})^2+
(\phi^1\bar\omega^{12})^2+(\bar\omega^{12}\omega_{23})^2\\&
+2(\bar\phi_3)^2\phi^1\omega_{23}+2(\phi^1)^2\bar\phi_3\bar\omega^{12}+
4\phi^1\bar\phi_3\bar\omega^{12}\omega_{23}\\&
+2\bar\phi_3(\omega_{23})^2\bar\omega^{12}+2\phi^1(\bar\omega^{12})^2\omega_{23}\,,
\end{array}
\label{t1}
\ee
\be
\begin{array}{rl}
\frac{(\theta_1)^2(\bar\theta^3)^2}{(\hat{\bar W}{}^{IJ}\hat W_{IJ})^2}=&
\frac{(\theta_1)^2(\bar\theta^3)^2}{(\bar\phi_I\phi^I)^2}
\left[ 1-2\frac{\phi^3\bar\omega^{12}}{\phi^I\bar\phi_I}
-2\frac{\bar\phi_1\omega_{23}}{\phi^I\bar\phi_I}
+3\frac{(\phi^3\bar\omega^{12})^2}{(\phi^I\bar\phi_I)^2}
+3\frac{(\bar\phi_1\omega_{23})^2}{(\phi^I\bar\phi_I)^2}\right.\\&
\left.
+6\frac{\phi^3\bar\phi_1\bar\omega^{12}\omega_{23}}{(\phi^I\bar\phi_I)^2}
-12\frac{(\phi^3\bar\omega^{12})^2\bar\phi_1\omega_{23}}{(\phi^I\bar\phi_I)^3}
-12\frac{\phi^3\bar\omega^{12}(\bar\phi_1\omega_{23})^2}{(\phi^I\bar\phi_I)^3}
+30\frac{(\phi^3\bar\omega^{12}\bar\phi_1\omega_{23})^2}{(\phi^I\bar\phi_I)^4}
\right].
\end{array}
\label{t2}
\ee
Substituting eqs. (\ref{t1}), (\ref{t2}) into the action (\ref{S1}) we
obtain
\be
\begin{array}{rl}
\hat S_{4}^{scale-inv}=&\alpha\dint d^4xd^{12}\theta \frac{(\theta_1)^2(\bar\theta^3)^2
(\bar\omega^{12}\omega_{23})^2}{(\phi^i\bar\phi_i)^2}\dint du\left[
30\frac{(\phi^1\bar\phi_1\phi^3\bar\phi_3)^2}{(\phi^i\bar\phi_i)^4}
\right.\\&
-\,24\dfrac{\phi^1\bar\phi_1(\phi^3\bar\phi_3)^2}{(\phi^i\bar\phi_i)^3}
-24\dfrac{(\phi^1\bar\phi_1)^2\phi^3\bar\phi_3}{(\phi^i\bar\phi_i)^3}
+3\dfrac{(\phi^3\bar\phi_3)^2}{(\phi^i\bar\phi_i)^2}
+3\dfrac{(\phi^1\bar\phi_1)^2}{(\phi^i\bar\phi_i)^2}\\&\left.
+24\,\dfrac{\phi^1\bar\phi_1\phi^3\bar\phi_3}{(\phi^i\bar\phi_i)^2}
-4\dfrac{\phi^3\bar\phi_3}{\phi^i\bar\phi_i}
-4\dfrac{\phi^1\bar\phi_1}{\phi^i\bar\phi_i}+1\right].
\end{array}
\label{S3}
\ee
The integrand in eq. (\ref{S3}) corresponds to the product of expressions (\ref{t1})
and (\ref{t2}) where only the terms with the maximal number of Grassmann variables
are considered since all other terms vanish under the integral.
The integration over the Grassmann variables in the action (\ref{S3}) yields
\be
\int d^{12}\theta
(\theta_1)^2(\bar\theta^3)^2(\bar\omega^{12}\omega_{23})^2
=16V^2\bar V^2
\ee
where we took into account the relation
\be
(\omega_{23}\bar\omega^{12})^2=16(\theta_2)^2(\theta_3)^2(\bar\theta^1)^2
(\bar\theta^2)^2 V^2\bar V^2.
\ee
To perform the integration over harmonic variables we apply the following
formula
\be
\int du (\phi^1\bar\phi_1)^m(\phi^3\bar\phi_3)^n=\frac{2m!n!}{(2+m+n)!}
(\phi^i\bar\phi_i)^{m+n}
\label{hint}
\ee
for each term in eq. (\ref{S3}). As a result, we obtain
\be
\hat S_4^{scale-inv}=\frac{\alpha_0}2\int d^4x\frac{V^2\bar V^2}{(\phi^i\bar\phi_i)^2}
\label{S4}
\ee
where $\alpha_0=\frac{32}{15}\alpha$.

Now we should express the auxiliary fields $V_{\alpha\beta},
\bar V_{\dot\alpha\dot\beta}$ through the physical field strengths
$F_{\alpha\beta}, \bar F_{\dot\alpha\dot\beta}$ from the action
\be
\hat S_2+\hat S_{4}^{scale-inv}=\int d^4x\left[V^2+\bar V^2-2(\bar V\bar F+ VF)+
\frac12(F^2+\bar F^2)+
\frac{\alpha_0}2\frac{V^2\bar V^2}{(\phi^i\bar\phi_i)^2}\right].
\label{S5}
\ee
The equations of motion for the auxiliary fields are
\be
2F_{\alpha\beta}=V_{\alpha\beta}\left[2+\frac{\alpha_0}{(\phi^i\bar\phi_i)^2}\bar
V^2\right],\qquad
2\bar F_{\dot\alpha\dot\beta}=\bar V_{\dot\alpha\dot\beta}
\left[2+\frac{\alpha_0}{(\phi^i\bar\phi_i)^2}V^2\right].
\label{EOM}
\ee
Eqs. (\ref{EOM}) define the auxiliary fields $V_{\alpha\beta},
\bar V_{\dot\alpha\dot\beta}$ as functions of
$F_{\alpha\beta}, \bar F_{\dot\alpha\dot\beta}$. The solution to these
equations can be represented as a series over the Maxwell field strengths:
\be
V_{\alpha\beta}=F_{\alpha\beta}\left[1-\frac{\alpha_0}{2(\phi^i\bar\phi_i)^2}
 \bar F^2+O(F^3)\right],\quad
\bar V_{\dot\alpha\dot\beta}=\bar F_{\dot\alpha\dot\beta}\left[1-
\frac{\alpha_0}{2(\phi^i\bar\phi_i)^2}F^2+O(F^3)\right].
\label{solEOM}
\ee
Substituting the solutions (\ref{solEOM}) into the action (\ref{S4}),
we find
\be
S_{4}^{scale-inv}=\frac{\alpha_0}2\int d^4x \left[\frac{F^2\bar F^2}{(\phi^i\bar\phi_i)^2}
-\frac{\alpha_0}2\frac{F^2\bar F^2}{(\phi^i\bar\phi_i)^4}(F^2+\bar F^2)
+O(F^8)
\right].
\label{S6}
\ee
Setting $\alpha_0=1$, we obtain the terms of 4-th and 6-th order in BI action
(\ref{BI}) with the correct coefficients.

Note that the action (\ref{S6}) contains all the higher-order terms starting with $F^8$.
However, as observed in \cite{Zupnik01,Ivanov02}, the coefficients in this series
are different from those in the field expansion of the BI action (\ref{BI}).
In the non-scale-invariant case this discrepancy is corrected
by the function $\hat E(A)$ in the action
(\ref{N3BI}).

Let us finish this Section with several further comments concerning the
superfield action (\ref{Seff}).

 This action contains the nonlocal operator
$\square^{-1}$.  However, the leading low-energy term in the component
action is local.  This can be explained as follows. Let us rewrite the
 action (\ref{Seff}) in the analytic superspace, using the relation
 (\ref{measures}).  We will obtain the lagrangian in which the
 derivatives $D^1_\alpha,\bar D_{3\dot\alpha}$ are distributed in all
 conceivable ways among the factors of the original lagrangian
 \p{Seff}. Consider one kind of such terms, namely those in which
 derivatives do not hit $\frac1{(\bar W^{IJ}W_{IJ})^2}$. Such terms
 either vanish due to the analyticity of superfield strengths $\bar
 W^{12}, W_{23}$, or result in local expressions, since the identities
\be
 (D^1)^2(\bar D_1)^2(W_{23})^2=-16\square (W_{23})^2\,, \quad
 (\bar D_3)^2(D^3)^2(\bar W^{12})^2=-16\square (\bar W^{12})^2
\ee
 allow one to cancel the factors $\square$ in the denominators of (\ref{Seff}).
 Precisely these superfield terms produce the component term $F^4/\phi^4$.
 Another type of terms corresponds to the situation when the derivatives $D^1_\alpha,\bar
 D_{3\dot\alpha}$ hit $\frac1{(\bar W^{IJ}W_{IJ})^2}$. It is easy to see that
 such terms produce the component expressions which contain the space-time
 derivatives of component fields and therefore these terms should be neglected
 in the low-energy approximation.

 From the very beginning there is a freedom in distributing the derivatives
 among different factors in the actions (\ref{S4full}) and (\ref{Seff}).
 However, the local part of the action (\ref{Seff}) actually does not depend
 on the specific pattern of such a distribution.
 The particular pattern we have chosen is most convenient for studying
 the component structure of the action. Let us dwell on this issue in more details.
 One can start with the action (\ref{S4full}) in
 which the derivatives $D^1_\alpha,\bar D_{3\dot\alpha}$ are distributed in
 a different way. Any such action can be cast in the form (\ref{S4full})
 by integrating by parts. But all such actions with diverse
 distributions of derivatives lead to different
 functionals of the type (\ref{Seff}) where the factor $\frac1{(\bar
 W^{IJ}W_{IJ})^2}$ is inserted. Integrating by parts, we see that
 all such actions differ only
 by non-local terms which appear when the derivatives hit the factor $\frac1{(\bar
 W^{IJ}W_{IJ})^2}$. As explained above, such terms are irrelevant for our consideration.

As follows from eq. (\ref{S6}), the action $S_4^{scale-inv}$
contains the term $\int d^4x F^4/(\phi^i\bar\phi_i)^2$ in its component
expansion. We observe that in this expression the scalar fields appear in a single
SU(3) invariant combination. An analogous result was earlier obtained in
ref. \cite{BuchIvanov,BuchIvanov_} for the full low-energy effective action of
$\cN=4$ SYM in the $\cN=2$ HSS approach.
The pivotal advantage of $\cN=3$ formalism consists in that all scalar fields
from the very beginning are included into a single
$\cN=3$ multiplet, while in the $\cN=2$ superspace language the scalar
fields are distributed between vector multiplet and hypermultiplet.

To summarize, the off-shell action (\ref{Seff}) is manifestly supersymmetric, gauge
invariant and scale invariant. It also bears the invariance under the
$\gamma_5$ and SU(3) transformations. Therefore, it can be considered as
a candidate for the low-energy effective action in $\cN=3$ SYM model.

\setcounter{equation}0
\section{Scale invariant $\cN=3$ BI action}

In the previous section we have demonstrated that the action (\ref{Seff}) is
responsible for the terms of 4-th and 6-th order in the scale invariant BI
action. Now we are going to construct $\cN=3$ superfield scale invariant generalization of
all other terms in BI action, starting with $F^8$.

A direct construction of scale invariant generalization of the action (\ref{N3BI})
in analytic superspace faces some difficulties. Therefore, in analogy
with the action (\ref{Seff}) we search for such a scale invariant
${\cal N}=3$ supersymmetric BI action in the full $\cN=3$ superspace. One of
possible superfield generalizations of higher terms in the scale invariant
BI action is provided by the following action
\be
S_G^{scale-inv}=\int d^4x
d^{12}\theta du \frac{(D^2)^2(\bar D_2)^2(\bar W^{IJ}W_{IJ})^2}{(\bar
W^{IJ}W_{IJ})^4} G\left(\frac{A}{(\bar W^{IJ}W_{IJ})^{4}}\right)
\label{snBMZ}
\ee
where $A$ is defined in eq. (\ref{A}) and
$ G$ is some function which can be represented as a series
\be
 G(a)=\sum^\infty_{n=0}\beta_n a^n
\label{ser1}
\ee
with some coefficients $\beta_n$.
The action (\ref{snBMZ}) is a scale invariant generalization of eq.
(\ref{N3BI}) in the sense that (as will be shown below)
it reproduces in components all terms in the scale invariant BI action, starting
from $F^8/\phi^{12}$, with definite coefficients
which can be fixed by choosing $\beta_{n}$ in the appropriate way.
Therefore, this action, taken in a sum with the quadratic $S_2$ and
quartic $S^{scale-inv}_4$ actions, can generate the scale invariant BI action in the bosonic
sector.

It should be noticed that the action (\ref{snBMZ}) is by no means the unique superfield expression
capable to reproduce the corresponding terms of the BI action
in the bosonic limit. There is a freedom in distributing derivatives
among different factors in eq. (\ref{snBMZ}). As we suppose, this freedom can be
compensated by the proper choice of function $G(a)$. Here we
consider just an example of such an action which is most convenient for
studying the component structure. We prove that a manifestly ${\cal N}=3$
supersymmetric scale invariant BI action exists, but do not discuss how unique
it is.\footnote{Perhaps, the freedom just mentioned could be fixed by requiring
the action to respect the full $\cN=3$ superconformal symmetry.}

When we consider only scalar and vector fields, the superfield strengths
are given by the ansatz (\ref{str-comp}). In such an approximation we derive
\be
\hat A=\frac1{2^{10}}(D^1)^2(\bar D_3)^2[
D^{2\alpha}\hat W_{12}D^2_\alpha\hat W_{12}\bar D_{2\dot\alpha}\hat{\bar W}{}^{23}
\bar D_2^{\dot\alpha}\hat{\bar
W}{}^{23}]=V^2\bar V^2,
\label{AA}
\ee
\be
(D^2)^2(\bar D_2)^2(\hat{\bar W}{}^{IJ}\hat W_{IJ})^2
=16^2V^2\bar V^2[(\bar\theta^1)^2(\theta_1)^2+(\bar\theta^3)^2(\theta_3)^2+
 2(\theta_1\theta_3)(\bar\theta^1\bar\theta^3)],
\label{b1}
\ee
\begin{eqnarray}
\dfrac{(\bar\theta^1)^2(\theta_1)^2}{(\hat{\bar W}{}^{IJ}\hat W_{IJ})^m}
&=&\dfrac{(\bar\theta^1)^2(\theta_1)^2}{(\bar\phi_i\phi^i)^{m+2}}
(\bar\omega^{23}\omega_{23})^2\frac{m(m+1)}{2}[1-2(m+2)\tfrac{\phi^1\bar\phi_1}{\phi^i\bar\phi_i}
\nonumber\\&&+\frac{(m+2)(m+3)}2\tfrac{(\phi^1\bar\phi_1)^2}{(\phi^i\bar\phi_i)^2}
]+\ldots,\label{bb2}\\
\dfrac{(\bar\theta^3)^2(\theta_3)^2}{(\hat{\bar W}{}^{IJ}\hat W_{IJ})^m}
&=&\dfrac{(\bar\theta^3)^2(\theta_3)^2}{(\bar\phi_i\phi^i)^{m+2}}
(\bar\omega^{12}\omega_{12})^2\frac{m(m+1)}{2}[1-2(m+2)\tfrac{\phi^3\bar\phi_3}{\phi^i\bar\phi_i}
\nonumber\\&&+\frac{(m+2)(m+3)}{2}\tfrac{(\phi^3\bar\phi_3)^2}{(\phi^i\bar\phi_i)^2}
]+\ldots,
\label{bb3}\\
\dfrac{(\theta_1\theta_3)(\bar\theta^1\bar\theta^3)}{(\hat{\bar
W}{}^{IJ}\hat W_{IJ})^m}&=&\dfrac{(\theta_1\theta_3)(\bar\theta^1\bar\theta^3)}{(\phi^i\bar\phi_i)^{m+2}}
\omega_{12}\omega_{23}\bar\omega^{12}\bar\omega^{23}m(m+1)
[1-(m+2)\tfrac{\phi^1\bar\phi_1+\phi^3\bar\phi_3}{\phi^i\bar\phi_i}\nonumber\\&&
+(m+2)(m+3)\tfrac{\phi^1\bar\phi_1\phi^3\bar\phi_3}{(\phi^i\bar\phi_i)^2}]+\ldots
\label{bb4}
\end{eqnarray}
where $m=4+4n$. Dots in eqs. (\ref{bb2})-(\ref{bb4}) correspond
to the terms with fewer number of Grassmann variables. These terms are not
important here since they do not contribute to the action
(\ref{snBMZ}).
Inserting the expressions (\ref{AA})-(\ref{bb4}) into the
action (\ref{snBMZ}) and performing there the integration over Grassmann and harmonic variables
with the help of eqs. (\ref{gme2}), (\ref{hint}), we find
\begin{eqnarray}
\hat S_{G}^{scale-inv}&=&
\sum^\infty_{n=0}\tilde\beta_n \int d^4x\frac{V^{2n+4}\bar V^{2n+4}}{
(\phi^i\bar\phi_i)^{6+4n}}
\label{snBMZcomp}
\end{eqnarray}
where $\tilde\beta_n=
2^{12}(1 + n)(5 + 4 n)(3 + 10 n + 8 n^2)\beta_{n}\,$.

The next steps are to express the auxiliary fields $V_{\alpha\beta},
\bar V_{\dot\alpha\dot\beta}$ through the field strengths $F_{\alpha\beta},
\bar F_{\dot\alpha\dot\beta}$ from the equations of motion generated by the
action
\be
\hat S_{BI}^{scale-inv}=\hat S_2+\hat S^{scale-inv}_4+\hat S^{scale-inv}_G
\label{BIcomp}
\ee
and to substitute these expressions back into the action (\ref{BIcomp}). As a
result, we should obtain the series over field strengths $F^2$, $\bar F^2$
with coefficients which depend on $\tilde\beta_n$. The coefficients
$\tilde\beta_n$ can be found from the requirement that the action
(\ref{BIcomp}), in the sector of vector and scalar fields, coincides with
the field expansion of the scale invariant version of the BI action (\ref{BI}).

For example, if we wish to fix the coefficient $\tilde\beta_0$ we should
expand the above action up to the terms $\sim F^8$:
\bea
\hat S_{BI}^{scale-inv} &=&
\int d^4x\left[V^2+\bar V^2-2(\bar V\bar F+ VF)+
\frac12(F^2+\bar F^2)+
\frac{1}2\frac{V^2\bar V^2}{(\phi^i\bar\phi_i)^2}
 \right. \nn \\
&&\left.  +\,\beta_0\frac{V^4\bar V^4}{(\phi^i\bar\phi_i)^6}
+ O(\frac{V^{6}\bar V^6)}{(\phi^i\bar\phi_i)^{10}}\right].
\label{BI-8}
\eea
The corresponding equations of motion for fields $V_{\alpha\beta}$, $\bar
V_{\dot\alpha\dot\beta}$  are
\begin{eqnarray}
\frac{\delta\hat S}{\delta V^{\alpha\beta}}&=&
2V_{\alpha\beta}-2F_{\alpha\beta}+\frac{V_{\alpha\beta}\bar V^2}{
(\phi^i\bar\phi_i)^2}+4\beta_0 V_{\alpha\beta}\frac{V^2\bar V^4}{
(\phi^i\bar\phi_i)^6}+\ldots=0\,,\label{eq1}\\
\frac{\delta\hat S}{\delta\bar V^{\dot\alpha\dot\beta}}&=&
2\bar V_{\dot\alpha\dot\beta}-2\bar F_{\dot\alpha\dot\beta}
+\frac{\bar V_{\dot\alpha\dot\beta} V^2}{
(\phi^i\bar\phi_i)^2}+4\beta_0\bar V_{\dot\alpha\dot\beta}
\frac{\bar V^2 V^4}{
(\phi^i\bar\phi_i)^6}+\ldots=0\,.\label{eq2}
\end{eqnarray}
The solutions to eqs. (\ref{eq1}), (\ref{eq2}) are given by the
following series
\begin{eqnarray}
V_{\alpha\beta}&=&
F_{\alpha\beta}\left(1-\frac12\frac{\bar F^2}{
(\phi^i\bar\phi_i)^2}+\frac12\frac{F^2\bar F^2}{
(\phi^i\bar\phi_i)^4}-\frac18\frac{
F^4\bar F^2+(4+16\beta_0)F^2\bar F^4}{
(\phi^i\bar\phi_i)^6}+\ldots\right),\label{eq3}\\
\bar V_{\dot\alpha\dot\beta}&=&
\bar F_{\dot\alpha\dot\beta}\left(1-\frac12\frac{F^2}{
(\phi^i\bar\phi_i)^2}+\frac12\frac{\bar F^2 F^2}{
(\phi^i\bar\phi_i)^4}-\frac18\frac{
\bar F^4 F^2+(4+16\beta_0)\bar F^2 F^4}{
(\phi^i\bar\phi_i)^6}+\ldots\right).\label{eq4}
\end{eqnarray}
Substituting the expressions (\ref{eq3}), (\ref{eq4}) back into the action
(\ref{BI-8}), we find that it coincides with the BI action (\ref{BI}) up to the
$F^8$ order, provided that $\tilde\beta_0=-\frac18$ and scalar fields are
substituted for the constant $X$, $X \rightarrow \phi^i\bar\phi_i$. The other coefficients
$\tilde\beta_n$ can be found in a similar way. Note that it is just the
technical problem to find the exact values for all these coefficients.

As a result, complete scale invariant $\cN=3$ BI action reads
\be
S_{BI}^{scale-inv}=S_2+S_4^{scale-inv}+S^{scale-inv}_{G}
\label{full}
\ee
where $S_2$ is the free quadratic action given by (\ref{S2_}), $S_4^{scale-inv}$ is
the non-local action (\ref{Seff}) corresponding to the $F^4/\phi^4$ term (as well as
yelding the 6th-order term) and $S^{scale-inv}_{G}$ given by (\ref{snBMZ}) is responsible for
all higher-order terms with correct coefficients.

To conclude this section, we briefly discuss the relationship between the
scale dependent BI action (\ref{N3BI}) and the scale invariant
BI action (\ref{snBMZ}). The first action is written as an integral over the analytic
subspace of $\cN=3$ superspace while the second one is given by an integral
over the full $\cN=3$ superspace. However, the
action (\ref{N3BI}) can also be rewritten as an integral over the full $\cN=3$ superspace.
The basic difference between them consists in the presence of
dimensional constant $X$ in eq. (\ref{N3BI}), while such a constant is absent in
the scale invariant action (\ref{snBMZ}). The above component analysis shows that
both these actions have the same component form in the bosonic sector,
provided one replaces the constant $X$ by $\phi^i\bar\phi_i$ and picks up the proper
functions $\hat E(a)$ in (\ref{N3BI}) and $G(a)$ in (\ref{snBMZ}).

\setcounter{equation}0
 \section{Summary}
 In this paper we analyzed the possible off-shell structure of
 low-energy effective action of $\cN=3$ SYM model written in
 $\cN=3$ harmonic superspace. This action was obtained as $\cN=3$ superfield
 generalization of the term $F^4/\phi^4$ which is leading in the low-energy
 effective action. This superfield action is written as a functional built out of
 the superfield strengths in full $\cN=3$ superspace. This functional is
 manifestly supersymmetric, gauge invariant, scale and $\gamma_5$-invariant and
 corresponds to a scale invariant generalization of 4-th order term in
 the $\cN=3$ supersymmetric BI action. We also constructed a
 scale invariant $\cN=3$ superfield completion of all higher terms in the BI action
 that might be helpful for studying higher-order terms in the candidate
 effective action of $\cN=3$ SYM model.

 Note that the expression (\ref{Seff}) is off-shell $\cN=3$ supersymmetric
 action. The superfield strengths $\bar W^{IJ}$, $W_{IJ}$ entering the
 action (\ref{Seff}) are unconstrained superfields.
 These superfields contain the scalar fields, gauge field strengths and fermions,
 all being combined into a single $\cN=3$ multiplet.
This means, in particular, that the on-shell $\cN=2$ vector multiplet and
hypermultiplet superfields from which the complete low-energy
$\cN=4$ SYM effective action is composed in the $\cN=2$ HSS \cite{BuchIvanov}
are now unified within a single off-shell $\cN=3$ gauge superfield.
 Therefore, the action (\ref{Seff}) can be regarded as an $\cN=3$
 superfield form of the complete $\cN=4$ supersymmetric low-energy
 effective action found in \cite{BuchIvanov}.

 In conclusion, let us point out once more that the effective action (\ref{Seff}) was
 found solely by employing the symmetries of the model and the requirement
 that it produces the $F^4/\phi^4$ term in components.
 This action was determined up to an arbitrary numerical coefficient.
  We would like to emphasize an analogy with the work
 \cite{Dine} where the on-shell low-energy effective action for
 $\cN=4$ SYM theory in the sector of $\cN=2$ gauge superfield was found merely
 on the symmetry grounds up to an arbitrary coefficient. This coefficient has
 been calculated later in the papers \cite{BuchN4} using the quantum
 field theory considerations. At present, $\cN=4$ SYM low-energy
 effective action is studied in many details both in the $\cN=2$
 gauge field sector (see e.g. \cite{BuchTseytlin}) and in the hypermultiplet
 sector \cite{BuchIvanov,BuchIvanov_}. The important role in direct calculations
 of the effective action in supersymmetric models is played by a superfield
 background field method (see e.g. \cite{book2,BK} for $\cN=1$ background field method
and \cite{backg} for $\cN=2$ background field method).
 It allows one to preserve manifest supersymmetry and gauge invariance at any step of quantum
 consideration. It is clear that the direct way of evaluating an effective
 action in $\cN=3$ supersymmetric gauge theory should be based upon
 a quantum $\cN =3$ formalism. The quantization procedure for this
 theory in $\cN = 3$ HSS was developed in \cite{Delduc}. The problem
 of working out the appropriate $\cN=3$ background field method becomes now
very important.

\section*{Acknowledgements}
 The work was partially supported by INTAS grant, project No 00-00254.
The work of I.L.B., E.A.I. and B.M.Z. was supported by DFG grant, project No 436
RUS 113/436 and RFBR-DFG grant, project No 02-02-04002.
 I.L.B. and I.B.S. are grateful to RFBR grant, project No
 03-02-16193, for partial support. The work of I.L.B. was supported by LSS
 grant, project No 1252.2003.2.  The work of I.B.S. was supported by LSS
 grant, project No 1743.2003.2.  The work of E.A.I. and B.M.Z. was
also supported by RFBR grant, project No 03-02-17440 and a grant of the Heisenberg-Landau
program.

\def\theequation{A.\arabic{equation}}
\setcounter{equation}0
\appendix
\section{Appendix}
 The $\cN=3$ HSS \cite{Harm2,IvanovBook} is defined as the superspace with coordinates $\{Z,u\}$, where
 $Z=\{x^{\alpha\dot\alpha},\theta_i^\alpha,\bar\theta^{i\dot\alpha} \}$
 \footnote{We denote by small Greek symbols the SL(2,$C$)
 spinor indices, $\alpha,\dot\alpha,\ldots=1,2$;
 the small Latin letters are SU(3) indices, $i,j,\ldots=1,2,3$.}
 is a set of standard $\cN=3$ coordinates and $u$ are the harmonics parameterizing
 the coset SU(3)$/$U(1)$\times$U(1). We consider the harmonics $u^I_i$ and their
 conjugate $\bar u_I^i$ ($I=1,2,3$) as SU(3) matrices
 \be
 u^I_i\bar u_J^i=\delta^I_J,\qquad
 u^I_i\bar u_I^j=\delta_i^j,\qquad
 \varepsilon^{ijk}u^1_iu^2_ju^3_k=1.
 \label{e1}
 \ee

 The harmonic superspace $\{Z,u\}$ contains the so called analytic subspace with
 the coordinates $\{\zeta_A,u\}=\{x_A^{\alpha\dot\alpha},\theta_2^\alpha,
 \theta_3^\alpha,\bar\theta^{1\dot\alpha},\bar\theta^{2\dot\alpha},u\}$ where
 \be
 x_A^{\alpha\dot\alpha}=x^{\alpha\dot\alpha}-
  2i(\theta_1^\alpha\bar\theta^{1\dot\alpha}-
  \theta_3^\alpha\bar\theta^{3\dot\alpha})\,, \qquad
 \theta_I^\alpha=\theta_i^\alpha \bar u^i_I\,, \qquad
 \bar\theta^{I\dot\alpha}=\bar\theta^{\dot\alpha i}u^I_i\,.
 \label{e2}
 \ee
 The analytic superspace plays an important role in harmonic superspace
 approach since it is closed under supersymmetry and all $\cN=3$ actions can be
 written in analytic coordinates.

 The harmonic superspace is equipped with Grassmann covariant derivatives
 $D^I_\alpha,\bar D_{I\dot\alpha}$ and harmonic covariant ones $D^I_J$ which
 form the su(3) algebra (see \cite{Harm2,IvanovBook} for details). The
 manifest expressions for these derivatives in the coordinates (\ref{e2})
 are
\be
\begin{array}{ll}
D^1_\alpha=\dfrac\partial{\partial\theta_1^\alpha}, &
 \bar D_{1\dot\alpha}=-\dfrac\partial{\partial\bar\theta^{1\dot\alpha}}-
   2i\theta_1^\alpha(\sigma^m)_{\alpha\dot\alpha}\partial_m,\\
D^2_\alpha=\dfrac\partial{\partial\theta_2^\alpha}+i\bar\theta^{2\dot\alpha}
  (\sigma^m)_{\alpha\dot\alpha}\partial_m,&
 \bar D_{2\dot\alpha}=-\dfrac\partial{\partial\bar\theta^{2\dot\alpha}}-
  i\theta_2^\alpha(\sigma^m)_{\alpha\dot\alpha}\partial_m,\\
D^3_\alpha=\dfrac\partial{\partial\theta_3^\alpha}+2i\bar\theta^{3\dot\alpha}
(\sigma^m)_{\alpha\dot\alpha}\partial_m,&
 \bar D_{3\dot\alpha}=-\dfrac\partial{\partial\bar\theta^{3\dot\alpha}},
\end{array}
\label{spinorderiv}
\ee
\be
\begin{array}l
D^1_2=\partial^1_2+
i\theta_2\sigma^m\bar\theta^1\partial_m+
 \bar\theta^{1\dot\alpha}\dfrac\partial{\partial\bar\theta^{2\dot\alpha}}-
 \theta_2^\alpha\dfrac\partial{\partial\theta_1^\alpha},\\
D^2_3=\partial^2_3+i\theta_3\sigma^m\theta^2\partial_m+
 \bar\theta^{2\dot\alpha}\dfrac\partial{\partial\bar\theta^{3\dot\alpha}}-
 \theta_3^\alpha\dfrac\partial{\partial\theta_2^\alpha},\\
D^1_3=\partial^1_3+2i\theta_3\sigma^m\theta^1
 \partial_m+
 \bar\theta^{1\dot\alpha}\dfrac\partial{\partial\bar\theta^{3\dot\alpha}}-
 \theta_3^\alpha\dfrac\partial{\partial\theta_1^\alpha},\\
D^2_1=\partial^2_1-i\theta_1\sigma^m\bar\theta^2
 \partial_m+
 \bar\theta^{2\dot\alpha}\dfrac\partial{\partial\bar\theta^{1\dot\alpha}}-
 \theta_1^\alpha\dfrac\partial{\partial\theta_2^\alpha},\\
D^3_2=\partial^3_2-i\theta_2\sigma^m\bar\theta^3\partial_m+
 \bar\theta^{3\dot\alpha}\dfrac\partial{\partial\bar\theta^{2\dot\alpha}}-
 \theta_2^\alpha\dfrac\partial{\partial\theta_3^\alpha},\\
D^3_1=\partial^3_1-2i\theta_1\sigma^m\bar\theta^3\partial_m+
 \bar\theta^{3\dot\alpha}\dfrac\partial{\partial\bar\theta^{1\dot\alpha}}-
 \theta_1^\alpha\dfrac\partial{\partial\theta_3^\alpha}
\end{array}
\label{harmderiv}
\ee
where $\partial^I_J=u^I_i\frac\partial{\partial u^J_i}-
\bar u_J^i\frac\partial{\partial\bar u_I^i}$.

The Grassmann and harmonic measures of integration over the $\cN=3$
analytic harmonic superspace are normalized so that
\begin{eqnarray}
\int d^{12}\theta\,
(\theta_1)^2(\theta_2)^2(\theta_3)^2(\bar\theta^1)^2(\bar\theta^2)^2
(\bar\theta^3)^2&=&1,\nonumber\\
\int
d\zeta(^{33}_{11})\,(\theta_2)^2(\theta_3)^2(\bar\theta^1)^2(\bar\theta^2)^2
&=&1,\nonumber\\
\int du&=&1.\label{gme2}
\end{eqnarray}

We use the following conventions for the relations between spinor and vector
indices
\be
x_{\alpha\dot\alpha}=(\sigma^m)_{\alpha\dot\alpha}x_m\,, \;\;
x_m=\frac12(\tilde\sigma_m)^{\dot\alpha\alpha}x_{\alpha\dot\alpha}\,.
\ee
Here $(\sigma^m)_{\alpha\dot\alpha}=(1,\vec\sigma)_{\alpha\dot\alpha}$ are Pauli matrices
and $(\tilde\sigma^m)^{\dot\alpha\alpha}=\varepsilon^{\alpha\beta}
\varepsilon^{\dot\alpha\dot\beta}(\sigma^m)_{\beta\dot\beta}\,$.
The corresponding relations for antisymmetric rank 2 tensor are
\bea
&& F_{mn}=\frac i2 F^{\alpha\beta}(\sigma_{mn})_{\alpha\beta}-
\frac i2\bar
F^{\dot\alpha\dot\beta}(\tilde\sigma_{mn})_{\dot\alpha\dot\beta}\,,\nn \\
&& F_{\alpha\beta}=\frac i4(\sigma_{mn})_{\alpha\beta}F^{mn},\quad
\bar F_{\dot\alpha\dot\beta}=-\frac
i4(\tilde\sigma_{mn})_{\dot\alpha\dot\beta}F^{mn}
\label{convert}
\eea
where
$$
(\sigma_{mn})_\alpha{}^\beta=\frac
i2(\sigma_m\tilde\sigma_n-\sigma_n\tilde\sigma_m)_\alpha{}^\beta\,,\qquad
(\tilde\sigma_{mn})^{\dot\alpha}{}_{\dot\beta}=\frac i2
(\tilde\sigma_m\sigma_n-\tilde\sigma_n\sigma_m)^{\dot\alpha}{}_{\dot\beta}\,.
$$

 \end{document}